\documentclass[12pt, draftclsnofoot, onecolumn]{IEEEtran}
\usepackage[cmex10]{amsmath}
\usepackage{amssymb}
\usepackage[utf8]{inputenc}
\usepackage[english]{babel}
\usepackage{amsthm}

\usepackage{array}
\usepackage[tight,footnotesize]{subfigure}
\usepackage[caption=false,font=footnotesize]{subfig}
\usepackage{cite}
\usepackage[usenames, dvipsnames]{color}
\usepackage[numbers,sort&compress]{natbib}
\usepackage{textcomp}
\usepackage{graphicx}
\usepackage{footnote}
\usepackage{epstopdf}
\usepackage{mathtools,amssymb,lipsum}

\makeatletter
\renewcommand{\fnum@figure}{Fig. \thefigure}
\makeatother

\usepackage{subfigure}
 
\begin{document}

\title{Error Performance of Various QAM Schemes for Nonregenerative Cooperative MIMO Network with Transmit Antenna Selection}
\author{Shaik Parvez, Praveen Kumar Singya, Vimal Bhatia, \IEEEmembership{Senior Member, IEEE} \\

		\IEEEauthorblockA{\small{Indian Institute of Technology Indore, India 453552.
			\\Email:	\{phd1601202003, phd1501102023, vbhatia\}@iiti.ac.in}}

}
\maketitle

\begin{abstract}
In this paper, performance of a dual-hop cooperative multiple-input and multiple-output (MIMO) system is analyzed over Rayleigh faded channels. For the considered system model, non-regenerative MIMO scheme with transmit antenna selection strategy (TAS) is employed. In this scheme, a single transmit antenna which maximizes the total received signal power at the receiver is selected for transmission.  Useful insights from average symbol error rate for various quadrature amplitude modulation (QAM) schemes such as general order hexagonal QAM (HQAM), general order rectangular QAM (RQAM) and 32-cross QAM (XQAM) are drawn. Monte-Carlo simulations are performed to validate the derived analytical expressions.   
\end{abstract}

\IEEEpeerreviewmaketitle
\section{Introduction}

		  Cooperative multiple-input and multiple-output (MIMO) communication systems has gained great interest due to the high data rates, improved spectral efficiency, enhanced coverage area and high diversity gain. Further, cooperative MIMO technology is preferred in IEEE 802.16m/j  and in LTE-Advanced, and is a promising technology for $5^{th}$ Generation (5G) mobile communication systems \cite{gupta2015survey}. In cooperative MIMO relay networks, transmit antenna strategy is employed  to reduce the system complexity, transmitter power and  hardware cost. It is also robust to channel estimation errors and require less feedback bits \cite{peters2008nonregenerative}. The amplify-and-forward (AF) MIMO cooperative relay networks due to their low complexity and easy in deployment have gained much research attention. \par 
		 In parallel, research on power and bandwidth efficient quadrature amplitude modulation (QAM) schemes has gained attention for future communication systems. The commonly used QAM signal constellations are rectangular QAM (RQAM), square QAM (SQAM) and cross QAM (XQAM). RQAM is considered as generic modulation scheme since it includes various modulation schemes \cite{kumar2018aser}. XQAM is an optimal constellation for transmission of odd number of bits per symbol because of low peak and average symbol energy than RQAM. Recently, HQAM scheme has received huge research interest due to its optimum two dimensional (2D) hexagonal lattice based constellation to achieve high data rates \cite{singya2018impact}. HQAM has low peak to average power ratio (PAPR) even for higher order constellations with optimum Euclidean distance and outperforms the other QAM schemes by providing considerable signal-to-noise ratio (SNR) gains. Thus cooperative MIMO network with QAM provides a viable solution for future communication systems.  \par{}
 	In \cite{singya2018impact,kumar2018aser} authors performed ASER analysis for HQAM, and RQAM for a single-input single-output (SISO) relay system. In \cite{peters2008nonregenerative}, \cite{cao2009transmit, duong2010performance, amarasuriya2010transmit, amarasuriya2011performance} transmit antenna selection (TAS) strategy is considered for the performance analysis of a MIMO relay system. In \cite{peters2008nonregenerative}, \cite{amarasuriya2010transmit, amarasuriya2011performance} multiple antennas are considered at all the respective nodes. Further, a direct link between the source and the destination is considered, and ASER analysis for binary phase shift keying (BPSK) is performed. In \cite{duong2010performance}, outage probability and the symbol error rate (SER) for M-ary phase shift keying (M-PSK) are derived for an AF-MIMO relay system without considering the direct link.

		   In this letter, for the first time, novel generalized closed-form expressions for the general order HQAM, general order RQAM and 32-XQAM are derived\ for a half-duplex  AF-MIMO relay network  with a direct link.

\section{System Model}
In this work, a dual-hop, half-duplex AF-MIMO relaying system model is considered where the source (S), relay (R) and the destination (D) are equipped with $N_{S}$, $N_{R}$, and $N_{D}$ antennas, respectively. It is assumed that the channel state information is available at both the relay and the destination nodes. Channel matrix from node $A$ to node $B$ is represented as $\textbf{H}_{AB}$, with dimensions $N_{r} \times N_{t}$, where $N_{r}$ and $N_{t}$ are the number of receiver and transmitter antennas respectively, and $A \in {\{S,R\}}$, $B \in {\{R,D\}}$, and $A \neq B$. Let $\textbf{h}_{AB}^{(i)} $ denote the $ N_{r}\times 1$ channel vector between the $i^{th}$ transmit antenna at node A and the $N_{r}$ receive antennas at node B of the channel matrix $\textbf{H}_{AB}$. All the channel gain vectors are assumed to be independent zero mean circularly symmetric Gaussian random variables with unit variance per dimension. In the first-hop, at S, a single transmit antenna  is selected to transmit the data to R and D nodes. The received signals at R ($\textbf{y}_{SR}$) and D ($\textbf{y}_{RD}$) can be written as 
\begin{align}
	\textbf{y}_{SR}&=\textbf{h}^{i}_{SR}x+\textbf{n}_{SR},	\\
	\textbf{y}_{SD}&=\textbf{h}^{i}_{SD}x+\textbf{n}_{SD},
	\end{align}
where  $ \textbf{h}^{i}_{SR} $ and $ \textbf{h}^{i}_{SD} $  are the channel gain vectors for  $S-R$ and $S-D$ links respectively, and $x$ is the transmitted signal with zero mean and unit variance. In the second hop, R amplifies the received signal and transmits the signal to D through the $k^{th}$ transmit antenna at R, which is expressed as
	\begin{align}
	\textbf{y}_{RD}&=\varphi(\textbf{h}^{k}_{RD}) (\textbf{h}^{i}_{SR})^{H} \textbf{y}_{SR}+\textbf{n}_{RD},
	\end{align}
where $\textbf{n}_{SR} $, $\textbf{n}_{SD} $, and $ \textbf{n}_{RD}$ are the  zero mean additive white Gaussian noise (AWGN) vectors associated with the $S-R$, $S-D$ and $R-D$ links, respectively and $\textbf{h}^{k}_{RD} $ is the channel gain vector of the $R-D$ link. $(\cdot)^H$ indicates the complex conjugate transpose. Further, $\varphi$ represents the amplification gain at the relay which is given as $\varphi \leq \frac{1}{\sqrt{{||h^{i}_{SR}||}^4+N_{0}{||h^{i}_{SR}||}^2}} \approxeq \frac{1}{{||h^{i}_{SR}||}^2}$ \cite{hasna2003end}. Here $||\cdot||^2$ indicates the squared Frobenius norm and $N_{0}$ indicates the variance of the AWGN. Finally, at D,  maximum ratio combining (MRC) is used to combine the signals received in two time slots with an minimum mean square error (MMSE) filter  \cite{peters2008nonregenerative},\cite{amarasuriya2010transmit}. Thus, the end-to-end (e2e) SNR at D is 
\begin{align} \lambda^{(i,k)}_{e2e}=	\lambda_{SD}^{(i)}+\frac{\lambda_{SR}^{(i)}\lambda_{RD}^{(k)}}{\lambda_{SR}^{(i)}+\lambda_{RD}^{(k)}},
\end{align}
where $\lambda_{AB}^{(j)}=\overline\lambda_{AB}||h_{AB}^j||^2$,  $\lambda_{AB}^{(j)}$ and $\overline{\lambda}_{AB}$ are the instantaneous and the average SNRs of the respective links. 
\section{Aser Analysis}
In this Section, the ASER expressions for various QAM schemes are derived by using the CDF approach. The generalized ASER expression for any digital modulation technique is given as  \cite[Eq. (5)]{kumar2018aser}
\begin{align}
P_{s}(e)=-\int_{0}^{\infty}P'_{s}(e|\lambda)F_{\lambda_{e2e}}(\lambda)d\lambda, 
\end{align}
where $P'_{s}(e|\lambda)$ represents the first order derivative of the conditional symbol error probability (SEP) with respect to (w.r.t) instantaneous SNR, $\lambda$, for the modulation scheme in AWGN channel, and $F_{\lambda_{e2e}}(\lambda)$ is the CDF of the received SNR \cite[Eq. (5)]{amarasuriya2010transmit} which is given as
\begin{align}
	P_{out}^{UB}(\lambda)=&F_{\lambda_{SD}}(\lambda)F_{\lambda_{SRD}}(\lambda), \\
	F_{\lambda_{SD}}(\lambda)= &\sum\limits_{v,w}^{} \mathbb{C}_{1}{{\lambda}^{w}}{{e}^{\frac{-v\lambda }{{{\lambda }_{0}}}}}, \nonumber\\ 
	F_{\lambda_{SRD}}(\lambda)=& 1+ \sum\limits_{i,j,m,n,r}\mathbb{C}_{2}\mathbb{C}_{3}{{e}^{-T\lambda }}{{\lambda }^{j+n+{{N}_{D}}}}{{K}_{\vartheta}}\Big( 2\sqrt{\chi}\lambda  \Big), \nonumber
\end{align}where $\sum\limits_{v,w}^{}=\sum\limits_{v=0}^{{{N}_{S}}} \sum\limits_{w=0}^{v ( {{N}_{D}}-1 )}$, $\mathbb{C}_{1}=\Big( \begin{matrix}
{{N}_{S}}  \\
v  \\
\end{matrix} \Big)\frac{{( -1 )}^{v} {\Omega }_{w,v,{{N}_{D}}}}{{\lambda_{0}}^{w}}$, $\sum\limits_{i,j,m,n,r}^{}=\sum\limits_{i=1}^{{{N}_{S}}}\sum\limits_{j=0}^{i( {{N}_{R}}-1 )}\sum\limits_{m=0}^{( {{N}_{R}}-1 )}\sum\limits_{n=0}^{m( {{N}_{D}}-1 )}\sum\limits_{r=0}^{j+n+{{N}_{D}}-1}$, $\mathbb{C}_{2}=(-1)^{i+m}\frac{2{{N}_{R}}{{\Omega }_{j,i,{{N}_{R}}}}{{\Omega }_{n,m,{{N}_{D}}}}}{\Gamma ( {{N}_{D}}){{\lambda }_{1}}^{\frac{r+j+1}{2}}{{\lambda }_{2}}^{\frac{2{{N}_{D}}+2n+j-r-1}{2}}}\frac{{{i}^{\frac{r-j+1}{2}}}}{{{( m+1 )}^{\frac{r+j+1}{2}}}}$, $\mathbb{C}_{3}=\Big( \begin{matrix}
{{N}_{S}}  \\
i  \\
\end{matrix} \Big)\Big( \begin{matrix}
{{N}_{R}}-1  \\
m  \\
\end{matrix} \Big)\Big( \begin{matrix}
j+n+{{N}_{D}}-1  \\
l  \\
\end{matrix} \Big)$, $\chi=\frac{i(m+1)}{{{\lambda }_{1}}{{\lambda }_{2}}} $, $ 
T=\frac{i}{{{\lambda }_{1}}}+\frac{m+1}{{{\lambda }_{2}}} $, $K_{\vartheta}(\cdot)$  is the modified Bessels function of second kind,  $\vartheta=r-j+1$, and $\Omega_{a,b,c}$ is the coefficient of the multinomial theorem $\Bigg(\sum\limits_{q=0}^{N-1} \frac{y^q}{q!}\Bigg)^r =\sum\limits_{p=0}^{r(N-1)}\Omega_{a,b,c} \ y^p $ \cite{amarasuriya2011performance}.
\begin{align}
&\Omega_{a,b,c}=\sum\limits_{i=a-c+1}^{a}\frac{\Omega_{i,b-1}}{(a-i)!} \ f_{[0,(b-1)(c-1)]}(i), \nonumber \\&
\Omega_{0,0,c}=\Omega_{0,b,c}=1, \  \Omega_{a,1,c}=\frac{1}{a!}, \ \Omega_{1,b,c}=b, \nonumber \\&
f_{[i,j]}(k)=\Bigg\{
\begin{array}{c l}	
1 & i<k<j \nonumber\\
0 & elsewhere
\end{array}\Bigg.
\end{align}
\subsection{Hexagonal QAM Scheme}
The conditional SEP expression for M-ary HQAM scheme over AWGN channel is given as \cite{rugini2016symbol}
\begin{align}
P_{s}(e|\lambda)&=MQ(\sqrt{\alpha \lambda})+\frac{2}{3}M_{c}Q^2\Bigg(\sqrt{\frac{2\alpha \lambda}{3}}\Bigg)-2M_cQ(\sqrt{\alpha \lambda})Q\Bigg(\sqrt{\frac{\alpha\lambda}{3}}\Bigg),	
\end{align}
where the numerical values of the parameters $M$, $M_c$ and $\alpha$ for different constellations are given in \cite{rugini2016symbol}. $Q(\cdot)$ is the Q-function denoted as $Q(x)=\frac{1}{\sqrt{2 \pi}}\int_{x}^{\infty} e^{\frac{-t^2}{2}}dt$. We derive the first order derivative of the conditional SEP expression (7) as 
\begin{align}
{{P}_{s}}^{'}\big( e|\lambda  \big)&=\frac{1}{2}\sqrt{\frac{\alpha }{2\pi }}\big( {{M}_{c}}-M \big){{\lambda }^{-\frac{1}{2}}}{{e}^{\frac{-\alpha \lambda }{2}}}-\frac{{{M}_{c}}}{3}\sqrt{\frac{\alpha }{3\pi }}{{\lambda }^{-\frac{1}{2}}}{{e}^{\frac{-\alpha \lambda }{3}}}+\frac{{{M}_{c}}}{2}\sqrt{\frac{\alpha }{6\pi }}{{\lambda }^{-\frac{1}{2}}}{{e}^{\frac{-\alpha \lambda }{6}}} \nonumber\\ 
& +\frac{2{{M}_{c}}}{9\pi }{{e}^{\frac{-2\alpha \lambda }{3}}}{}_{1}{{F}_{1}}\Big( 1,\frac{3}{2},\frac{\alpha \lambda }{3} \Big)-\frac{{{M}_{c}}\alpha }{2\sqrt{3}\pi }{{e}^{\frac{-2\alpha \lambda }{3}}}\Big\{{}_{1}{{F}_{1}}\Big( 1,\frac{3}{2},\frac{\alpha \lambda }{2} \Big)+{}_{1}{{F}_{1}}\Big( 1,\frac{3}{2},\frac{\alpha }{6}\lambda  \Big) \Big\} 
\end{align}
 where ${}_{1}{{F}_{1}}(a,b,c)$ is the confluent hypergeometric function. Further, substituting $P_s^{'}(e|\lambda)$ and $F_{\lambda_{e2e}}(\lambda)$ from (8) and (6), respectively into (5),   and by solving the integrals using the following identities \cite[eq. (3.371), (7.522.9), (6.621.3)]{gradshteyn2014table} and \cite[eq. (8)]{romero2009performance}, the closed-form ASER expression for the HQAM scheme can be derived as 
\begin{align}
  P_s^{HQAM}(e)&=I_{H_{1}}(e)+I_{H_{2}}(e), \nonumber\\
{{I}_{H_{1}}}(e)&=\Xi1\Big( \Im_{1}
+h_{1}\Im_{2}
-\Im_{3}\Big),  \\
{{I}_{H_{2}}}(e)&=\Xi2\Big(h_{2} \big( 
\mathbb{A}_{1}A_{1} -\mathbb{A}_{2} A_{2} + \mathbb{A}_{3}A_{3} \big) + \Sigma \ h_{3} \mathbb{A}_{4}\Big),
\end{align}
where $\Xi1 =-\sum\limits_{v,w}^{} \mathbb{C}_{1}$, $\Xi2=-\sum\limits_{i,j,m,n,r}\sum\limits_{v,w}^{} \mathbb{C}_{1}\mathbb{C}_{2}\mathbb{C}_{3}$. Further,  $\Im_{1}=( w-\frac{1}{2})!\big( \frac{1}{2}\sqrt{\frac{\alpha }{2\pi }}\big( {{M}_{c}}-M \big){{\big( \frac{\alpha }{2}+\frac{v}{{{\lambda }_{0}}} \big)}^{-\theta}}-\frac{{{M}_{c}}}{3}\sqrt{\frac{\alpha }{3\pi }}{{\big( \frac{\alpha }{3}+\frac{v}{{{\lambda }_{0}}} \big)}^{-\theta}}+\frac{{{M}_{c}}}{2}\sqrt{\frac{\alpha }{6\pi }}{{\big( \frac{\alpha }{6}+\frac{v}{{{\lambda }_{0}}} \big)}^{-\theta}} \big)$, $\theta= w+\frac{1}{2} $, $h_{1}=\Gamma(\theta_{1}) {{\big( \frac{2\alpha }{3}+\frac{v}{{{\lambda }_{0}}} \big)}^{-\theta_{1}}}$, $\Im_{2}= \frac{2{{M}_{c}}\alpha }{9\pi }{}_{2}{{F}_{1}}\big( 1,\theta_{1},\frac{3}{2},\frac{\frac{\alpha }{3}}{\frac{2\alpha }{3}+{{\frac{v}{{{\lambda }_{0}}} }}} \big) $,
$\Im_{3}= \frac{{{M}_{c}}\alpha }{2\sqrt{3}\pi }{}_{2}{{F}_{1}}\big( 1,\theta_{1},\frac{3}{2},\frac{\frac{\alpha }{2}}{\frac{2\alpha }{3}+{{ \frac{v}{{{\lambda }_{0}}} }}} \big)+{}_{2}{{F}_{1}}\big( 1,\theta_{1},\frac{3}{2},\frac{\frac{\alpha }{6}}{\frac{2\alpha }{3}+{{\frac{v}{{{\lambda }_{0}}}}}} \big)$, $\theta_{1}=w+1$, $h_{2}=\frac{\sqrt{\pi }( 2{{\beta }_{1}} )\Gamma ({{u}_{1}}+{\vartheta})\Gamma ({{u}_{1}}-{\vartheta})}{\Gamma ({{u}_{1}}+\frac{1}{2})}$,
$\mathbb{A}_{1}=\frac{1}{2}\sqrt{\frac{\alpha }{2\pi }}({{M}_{c}}-M)$, $A_{1}=\frac{1}{{{( {{\alpha }_{1}}+{{\beta }_{1}} )}^{{{u}_{1}}+{\vartheta}}}}{}_{2}{{F}_{1}}\big( {{u}_{1}}+{\vartheta},{\vartheta}+\frac{1}{2},{{u}_{1}}+\frac{1}{2},\frac{{{\alpha }_{1}}+{{\beta }_{1}}}{{{\alpha }_{1}}-{{\beta }_{1}}} \big)$, $\mathbb{A}_{2}=\frac{{{M}_{c}}}{3}\sqrt{\frac{\alpha }{3\pi }}$
$A_{2}=\frac{1}{{{( {{\alpha }_{2}}+{{\beta }_{1}} )}^{{{u}_{1}}+{\vartheta}}}}{}_{2}{{F}_{1}}\big( {{u}_{1}}+{\vartheta},{\vartheta}+\frac{1}{2},{{u}_{1}}+\frac{1}{2},\frac{{{\alpha }_{2}}+{{\beta }_{1}}}{{{\alpha }_{2}}-{{\beta }_{1}}} \big) $,
$\mathbb{A}_{3}=\frac{{{M}_{c}}}{2}\sqrt{\frac{\alpha }{6\pi }}$,
$A_{3}=\frac{1}{{{( {{\alpha }_{3}}+{{\beta }_{1}} )}^{{{u}_{1}}+{\vartheta}}}}{}_{2}{{F}_{1}}\big( {{u}_{1}}+{\vartheta},{\vartheta}+\frac{1}{2},{{u}_{1}}+\frac{1}{2},\frac{{{\alpha }_{3}}+{{\beta }_{1}}}{{{\alpha }_{3}}-{{\beta }_{1}}} \big)$,
$\Sigma=\sum\limits_{z=0}^{\infty }{\frac{{{( 1 )}_{z}}}{{{( \frac{3}{2} )}_{z}}z!}}$,
$h_{3}=\frac{\sqrt{\pi }( 2{{\beta }_{1}} )}{{{( {{\alpha }_{4}}+{{\beta }_{1}} )}^{{{u}_{2}}+{\vartheta}}}}\frac{\Gamma ({{u}_{2}}+{\vartheta})\Gamma ({{u}_{2}}-{\vartheta})}{\Gamma ({{u}_{2}}+\frac{1}{2})}{}_{2}{{F}_{1}}\Big( {{u}_{2}}+{\vartheta},{\vartheta}+\frac{1}{2},{{u}_{2}}+\frac{1}{2},\frac{{{\alpha }_{4}}+{{\beta }_{1}}}{{{\alpha }_{4}}-{{\beta }_{1}}} \Big)$,
$\mathbb{A}_{4}=\Big( \frac{{{-M}_{c}}\alpha }{2\sqrt{3}\pi }\big( {{\big( \frac{\alpha }{2} \big)}^{z}}+{{\big( \frac{\alpha }{6} \big)}^{z}} \big)+\frac{2{{M}_{c}}\alpha }{9\pi }{{\big( \frac{\alpha }{3} \big)}^{z}} \Big)$, ${{u}_{1}}=j+n+{{N}_{D}}+w+\frac{1}{2} $, $ {{u}_{2}}=j+n+{{N}_{D}}+w+z+1 $, ${{\beta }_{1}}=2\sqrt{\chi}$,
 ${{\alpha }_{1}}=T+\frac{\alpha }{2}+\frac{v}{{{\lambda }_{0}}}$, ${{\alpha }_{2}}=T+\frac{\alpha }{3}+\frac{v}{{{\lambda }_{0}}}$, ${{\alpha }_{3}}=T+\frac{\alpha }{6}+\frac{v}{{{\lambda }_{0}}}$, $ {{\alpha }_{4}}=T+\frac{2\alpha }{3}+\frac{v}{{{\lambda }_{0}}}$, 
${}_{2}{{F}_{1}}(a,b,c,d)$ is a hypergeometric function, and $(a)_{b}$ is the Pochhammer symbol which is $(a)_{b}=\frac{\Gamma(a+b)}{\Gamma(a)}$.
\subsection{Rectangular QAM Scheme}
ASER expression for the RQAM scheme is also derived by using the CDF approach by considering the first order derivative of the conditional SEP w.r.t. $\lambda$. Over the AWGN channel, the conditional SEP for RQAM scheme is given as \cite[eq. (14)]{singya2018impact}
\begin{align}
{{P}_{s}}^{RQAM}(e|\lambda )& =2\Big[ {\mathbb{N}_{1}}Q\big( \zeta \sqrt{\lambda } \big)+{\mathbb{N}_{2}}Q\big( \rho \sqrt{\lambda } \big) -2{\mathbb{N}_{1}}{\mathbb{N}_{2}}Q\big( \zeta \sqrt{\lambda } \big)Q\big( \rho \sqrt{\lambda } \big) \Big],
\end{align}
where ${\mathbb{N}_{1}}=1-\frac{1}{{{M}_{I}}}$,  ${\mathbb{N}_{2}}=1-\frac{1}{{{M}_{Q}}}$, $\zeta =\sqrt{\frac{6}{( {{M}_{I}}^{2}-1 )+( M_{Q}^{2}-1){{\sigma }^{2}}}} $, $\rho =\sigma \zeta $,  wherein $M_{I}$ and $M_{Q}$ are the number of in-phase and quadrature-phase constellation points respectively. Also, $\sigma =\frac{dQ}{dI}$, where $d_{I}$ and $d_{Q}$ indicate the in-phase and quadrature decision distances respectively. First order derivative of the conditional SEP can be derived by using the aforementioned identities as
\begin{align}
P_{RQAM}^{'}(e|\lambda )=&\frac{\zeta{\mathbb{N}_{1}}( {\mathbb{N}_{2}}-1 )}{\sqrt{2\pi \lambda }}{{e}^{\frac{-{{\zeta }^{2}}}{2}\lambda }}+\frac{\rho {\mathbb{N}_{2}}( {\mathbb{N}_{1}}-1)}{\sqrt{2\pi \lambda }}{{e}^{\frac{-{{\rho }^{2}}}{2}\lambda }}-\nonumber\\&\frac{\zeta \rho {\mathbb{N}_{1}}{\mathbb{N}_{2}}}{\pi}\big\{ {}_{1}{{F}_{1}}\big( 1,\frac{3}{2},\frac{{{\rho }^{2}}}{2}\lambda  \big)+{}_{1}{{F}_{1}}\big( 1,\frac{3}{2},\frac{{{\zeta }^{2}}}{2}\lambda  \big) \big\}{{e}^{\frac{-( {{\zeta }^{2}}+{{\rho }^{2}} )\lambda }{2}}}.
\end{align}
 Further, substituting $P_s^{'}(e|\lambda)$ and $F_{\lambda_{e2e}}(\lambda)$, respectively into (5), and by solving the integrals using following identities \cite[eq. (3.371), (7.522.9), (6.621.3)]{gradshteyn2014table} and \cite[eq. (8)]{romero2009performance}, closed-form ASER expression for the RQAM scheme can be derived as
\begin{align}
&P_s^{RQAM}(e)=I_{R_{1}}(e)+I_{R_{2}}(e),  \\
&{{I}_{R_{1}}}(e)=\Xi1\big(B_{1}( \Bbbk_{1}+\Bbbk_{2}) -B_{2}(\Bbbk_{3}+\Bbbk_{4}) \big), \\
&{{I}_{R_{2}}}(e)=\Xi2\Big(B_{3}( \Bbbk_{5}+\Bbbk_{6} )  +\Sigma \ B_{4} \ \Bbbk_{7}\Bbbk_{8}\Big),
\end{align}
where  $B_{1}={2}^{-\theta}( 2w-1 )!!$, $\Bbbk_{1}={{\zeta {{g}_{1}}( {{g}_{2}}-1 )\big( \frac{{\zeta}^{2}}{2}+\frac{v}{{{\lambda }_{0}}} \big)}^{-\theta}}$, $\Bbbk_{2}={{\rho {{g}_{2}}( {{g}_{1}}-1 )\big( \frac{{\rho}^{2}}{2}+\frac{v}{{{\lambda }_{0}}} \big)}^{-\theta}}$, $B_{2}=\frac{{\zeta}{\rho}{{g}_{2}}{{g}_{1}}}{\pi }\Gamma (\theta_{1} ) {{\big(\frac{{\zeta}^{2}}{2}+\frac{{\rho}^{2}}{2}+\frac{v}{{{\lambda }_{0}}} \big)}^{-\theta_{1}}}$, $\Bbbk_{3}={}_{2}{{F}_{1}}\big( 1,\theta_{1},\frac{3}{2},\frac{\frac{{
			\rho}^{2}}{2}}{\frac{{\zeta}^{2}}{2}+\frac{{
			\rho}^{2}}{2}+\frac{v}{{{\lambda }_{0}}}} \big)$, $\Bbbk_{4}={}_{2}{{F}_{1}}\big( 1,\theta_{1},\frac{3}{2},\frac{\frac{{\zeta}^{2}}{2}}{\frac{{\zeta}^{2}}{2}+\frac{{
			\rho}^{2}}{2}+\frac{v}{{{\lambda }_{0}}}} \big)$, $\theta= w+0.5 $, $\theta_{1}=w+1$, 
		$B_{3}=\frac{\sqrt{\pi }{{( 2{{\beta }_{1}})}^{{\vartheta}}}}{\sqrt{2\pi }}\frac{\Gamma ({{u}_{1}}+{\vartheta})\Gamma ({{u}_{1}}-{\vartheta})}{\Gamma ({{u}_{1}}+0.5)}$
		$\Bbbk_{5}=\frac{{\zeta}{{g}_{1}}( {{g}_{2}}-1 )}{{{( {{\alpha }_{1}}+{{\beta }_{1}} )}^{{{u}_{1}}+{\vartheta}}}}
		{}_{2}{{F}_{1}}\big( {{u}_{1}}+{\vartheta},{\vartheta}+0.5,{{u}_{1}}+0.5,\frac{{{\alpha }_{1}}+{{\beta }_{1}}}{{{\alpha }_{1}}-{{\beta }_{1}}} \big)$
		$\Bbbk_{6}= \frac{{
				\rho}{{g}_{2}}( {{g}_{1}}-1 )}{{{( {{\alpha }_{2}}+{{\beta }_{1}} )}^{{{u}_{1}}+{\vartheta}}}}{}_{2}{{F}_{1}}\big( {{u}_{1}}+{\vartheta},{\vartheta}+0.5,{{u}_{1}}+0.5,\frac{{{\alpha }_{2}}+{{\beta }_{1}}}{{{\alpha }_{2}}-{{\beta }_{1}}} \big)$
		$B_{4}=\frac{-{\zeta}{
				\rho}{{g}_{2}}{{g}_{1}}}{\pi}\frac{\sqrt{\pi }{{( 2{{\beta }_{1}})}^{{\vartheta}}}}{{{( {{\alpha }_{3}}+{{\beta }_{1}})}^{{{u}_{2}}+{\vartheta}}}}\frac{\Gamma ({{u}_{2}}+{\vartheta})\Gamma ({{u}_{2}}-{\vartheta})}{\Gamma ({{u}_{2}}+0.5)}$
		$\Bbbk_{7}={}_{2}{{F}_{1}}\big( {{u}_{2}}+{\vartheta},{\vartheta}+0.5,{{u}_{2}}+0.5,\frac{{{\alpha }_{3}}+{{\beta }_{1}}}{{{\alpha }_{3}}-{{\beta }_{1}}} \big)$, and $\Bbbk_{8}=\big( {{\big( \frac{{\zeta}^{2}}{2} \big)}^{_{z}}}+{{\big( \frac{{
						\rho}^{2}}{2} \big)}^{_{z}}} \big)$. Further, $ {{u}_{1}}=j+n+{{N}_{D}}+w+0.5$, ${{u}_{2}}=j+n+{{N}_{D}}+w+z+1$, ${{\alpha }_{1}}=T+\frac{v}{{{\lambda }_{0}}}+\frac{{\zeta}^{2}}{2}$, ${{\beta }_{1}}=2\sqrt{\chi} $, ${{\alpha }_{2}}=T+\frac{v}{{{\lambda }_{0}}}+\frac{{
		\rho}^{2}}{2}$, ${{\alpha }_{3}}=T+\frac{v}{{{\lambda }_{0}}}+\frac{{\zeta}^{2}+{
		\rho}^{2}}{2}$.
\subsection{Cross QAM Scheme}
The conditional SEP for the 32-XQAM scheme is given as \cite[eq. (21)]{zhang2010exact}
\begin{align}
{{P}_{s}}^{XQAM}\big( e|\lambda  \big)&={{E}_{1}}Q\big( \sqrt{2C\lambda } \big)+\frac{4}{M}Q\big( 2\sqrt{C\lambda } \big)-{{E}_{2}}{{Q}^{2}}\big( \sqrt{2C\lambda } \big),
\end{align}
where $E_{1}=4-\frac{6}{\sqrt{2M}}$, $E_{2}=4-\frac{12}{\sqrt{2M}}+\frac{12}{M}$, $C=\frac{48}{31M-32}$, and $M=32$. 
The first order derivative of conditional SEP expression for 32-XQAM is given as
\begin{align}
P_{XQAM}^{'}\big( e|\lambda  \big)=&\frac{-{{E}_{1}}}{2}\sqrt{\frac{C}{\pi }}{{e}^{-C\lambda }}{{\lambda }^{-0.5}}-\frac{4}{M}\sqrt{\frac{C}{2\pi }}{{e}^{-2C\lambda }}{{\lambda }^{-0.5}}+\frac{{{E}_{2}}}{2}\sqrt{\frac{C}{\pi }}{{e}^{-C\lambda }}{{\lambda }^{-0.5}} \nonumber \\  &
-\frac{{{E}_{2}}C}{\pi }{}_{1}{{F}_{1}}( 1,\frac{3}{2},C\lambda ){{e}^{-2C\lambda }}.
\end{align} 
 The closed-form ASER expression for 32-XQAM scheme can be derived by substituting $ P_{XQAM}^{'}\big( e|\lambda  \big)$ and $F_{\lambda_{e2e}}(\lambda)$ in (5). The integrals are resolved by using the identities \cite[eq. (3.371), (7.522.9), (6.621.3)]{gradshteyn2014table} and \cite[eq. (8)]{romero2009performance}. The final closed-form ASER expression for 32-XQAM is 
 \begin{align}
 P_s^{XQAM}(e)&=I_{X_{1}}(e)+I_{X_{2}}(e), \\
 {{I}_{X_{1}}}( e )&=\Xi1\Big({2}^{-v}\big( 2v-1 \big)!!\big(\omega_{1}+\omega_{2}\big)-\Gamma (v+1)\omega_{3}\Big),	\\
 {{I}_{X_{2}}}( e )&= \Xi2 \Big( \Theta_{1} \big(\omega_{4}\big( \Theta_{2}+\Theta_{3} \big)+\Theta_{4}\omega_
 {5}\big) 	+ \Sigma \Theta_{5} \omega_{6}  \Big),
 \end{align}
where $\omega_{1}={{\big( C+\frac{u}{{{\lambda }_{0}}} \big)}^{-( v+\frac{1}{2} )}}\Big( \frac{-{{E}_{1}}}{2}\sqrt{C}+\frac{{{E}_{2}}}{2}\sqrt{C} \Big)$, $\omega_{2}=\frac{-4}{M}\sqrt{\frac{C}{2}}{{\Big( 2C+\frac{u}{{{\lambda }_{0}}} \Big)}^{-( v+\frac{1}{2} )}}$, \\ \nonumber
$\omega_{3}=\frac{{{E}_{2}}C}{\pi }{{\big( 2C+\frac{u}{{{\lambda }_{0}}} \big)}^{-( v+\frac{1}{2} )} {}_{2}{{F}_{1}}\big( 1,v+1,\frac{3}{2},\frac{C}{2C+\frac{u}{{{\lambda }_{0}}}} \big)}$, $\Theta_{1}=\frac{\sqrt{\pi }{{( 2{{\beta }_{1}} )}^{{\vartheta}}}\Gamma ( {{u}_{1}}+{\vartheta} )\Gamma ( {{u}_{1}}-{\vartheta} )}{\Gamma ( {{u}_{1}}+\frac{1}{2} \big)}$,
$\omega_{4}={}_{2}{{F}_{1}}\big( {{u}_{1}}+{\vartheta},{\vartheta}+\frac{1}{2},{{u}_{1}}+\frac{1}{2},\frac{{{\alpha }_{1}}-{{\beta }_{1}}}{{{\alpha }_{1}}+{{\beta }_{1}}} \big)$,
$\Theta_{2}=\frac{\frac{-{{E}_{1}}}{2}\sqrt{\frac{C}{\pi }}}{{{( {{\alpha }_{1}}+{{\beta }_{1}} )}^{{{u}_{1}}+{\vartheta}}}}$,
$\Theta_{3}=\frac{\frac{{{E}_{2}}}{2}\sqrt{\frac{C}{\pi }}}{{{( {{\alpha }_{1}}+{{\beta }_{1}} )}^{{{u}_{1}}+{\vartheta}}}}$
$\Theta_{4}=\frac{\frac{-4}{M}\sqrt{\frac{C}{2\pi }}}{{{( {{\alpha }_{2}}+{{\beta }_{1}} )}^{{{u}_{1}}+{\vartheta}}}}$,
$\omega_{5}={}_{2}{{F}_{1}}\big( {{u}_{1}}+{\vartheta},{\vartheta}+\frac{1}{2},{{u}_{1}}+\frac{1}{2},\frac{{{\alpha }_{2}}-{{\beta }_{1}}}{{{\alpha }_{2}}+{{\beta }_{1}}} \big)$, $\Theta_{5}=\frac{-{{E}_{2}}C}{\pi }\frac{\sqrt{\pi }{{( 2{{\beta }_{1}} )}^{{\vartheta}}}}{{{( {{\alpha }_{2}}+{{\beta }_{1}} )}^{{{u}_{1}}+{\vartheta}}}}\frac{\Gamma ( {{u}_{2}}+{\vartheta} )\Gamma ( {{u}_{2}}-{\vartheta} )}{\Gamma ( {{u}_{2}}+\frac{1}{2} )}$, and
$\omega_{6}={}_{2}{{F}_{1}}\big( {{u}_{2}}+{\vartheta},{\vartheta}+\frac{1}{2},{{u}_{2}}+\frac{1}{2},\frac{{{\alpha }_{2}}-{{\beta }_{1}}}{{{\alpha }_{2}}+{{\beta }_{1}}} \big)$. Further, ${{u}_{1}}=j+n+{{N}_{D}}+w+\frac{1}{2}$, ${{u}_{2}}=j+n+{{N}_{D}}+w+z+1$, ${{\beta }_{1}}=2\sqrt{\chi} $, ${{\alpha }_{1}}=T+\frac{u}{{{\lambda }_{0}}}+C$, and ${{\alpha }_{2}}=T+\frac{u}{{{\lambda }_{0}}}+2C$.

\begin{figure}
	\centering
	\includegraphics[width=4in,height=3in]{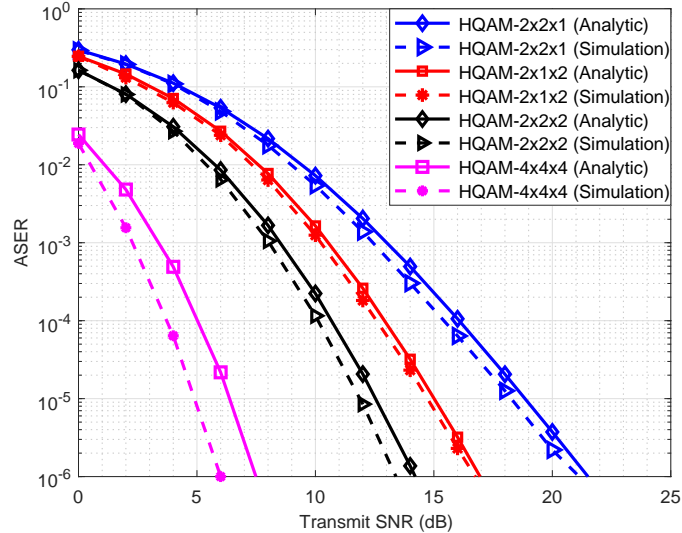}
	\caption{ ASER for 8-HQAM scheme with different antenna configurations.}
\end{figure}
\begin{figure*}[htp]
	\centering
	\subfigure[2x2x2 antenna configuration.]{\includegraphics[width=4in,height=3in]{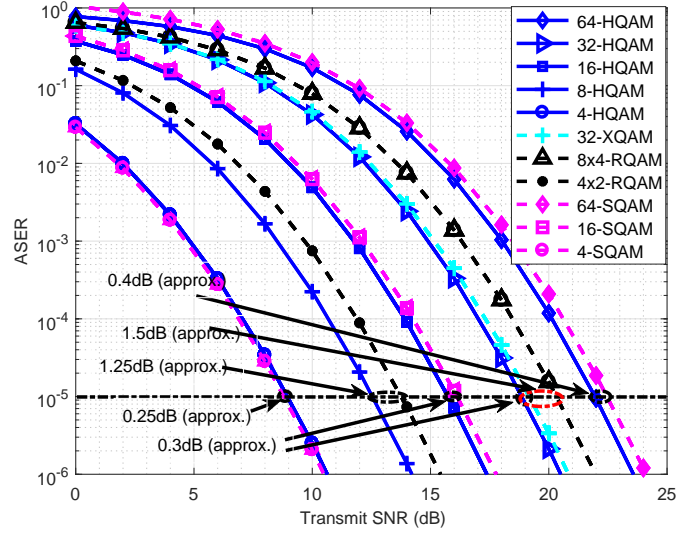}}\quad
	\subfigure[4x4x4 antenna configuration.]{\includegraphics[width=4in,height=3in]{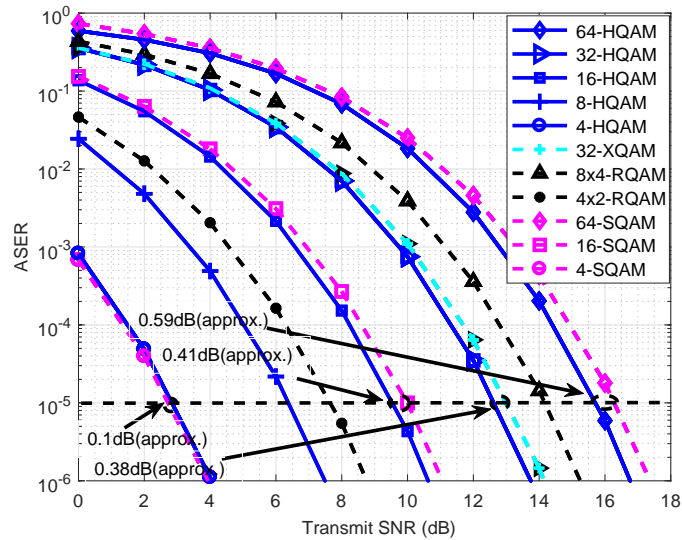}}\quad
	\caption{Comparison of ASER results of various QAM schemes for different MIMO antenna configurations.}
\end{figure*}

\section{Numerical and Simulation Results}

	In this Section, simulation results are obtained to validate the derived analytical expressions. Analytical expressions are numerically evaluated for the considered system model. The impact of the network geometry is considered by modeling the average SNR as $\overline{\lambda}_{AB} = \overline{\lambda}_{SD} (\frac{D_{SD}}{D_{AB}})^\phi$, where $\overline{\lambda}_{SD}$ is the average SNR of the direct link and $\phi$ is the pathloss exponent and is considered to be $2.5$. $D_{SD}$ is the distance from source to destination and, $D_{SR}=\frac{D_{SD}}{3}$ and $D_{RD}=\frac{2D_{SD}}{3}$ are the S-R and R-D distances, respectively \cite{amarasuriya2010transmit}. \par
	In Fig. 1, analytical and simulation results of ASER for 8-HQAM scheme for different MIMO  antenna configurations are presented when direct path is considered. From Fig. 1, it is observed that the theoretical curves are always above the simulation curves which validates the upper-bound of the derived ASER results. Further, the simulation results are close to the theoretical results which validates the derived analytical expressions. For 2x2x2 and 4x4x4 antenna configurations, the results are bounded within 1 dB and about 1.5 dB respectively, similar observations are made in \cite{amarasuriya2010transmit}. \par
	In Fig. 2(a) and Fig. 2(b), analytical ASER results for the HQAM are compared with SQAM, XQAM, and RQAM schemes for various constellation points for 2x2x2 and 4x4x4 antenna configurations, respectively. It is evident that the HQAM outperforms the other schemes except for 4-HQAM, since 4-HQAM has larger average number of nearest neighbors (M) than 4-SQAM \cite{rugini2016symbol}. From Fig. 2(a), it is observed that for ASER of $10^{-5} $,  4-SQAM performs better than 4-HQAM with an SNR gain of 0.25 dB (approx.). However, for higher order QAM constellations HQAM outperforms the other QAM schemes. 16 HQAM has a SNR gain of 0.3 dB over 16-SQAM which further increases to 0.4 dB for 64-HQAM over 64-SQAM. For 8 constellation points, HQAM provides an SNR gain of around 1.25 dB over 4x2-RQAM scheme. For 32 points, 32-XQAM has an SNR gain of 1.2 dB over 8x4-RQAM due to the lower peak and average energy of XQAM constellation than RQAM. HQAM has an SNR gain of 0.3 dB over 32-XQAM. This is due to the increase in $ \alpha $ value with constellation points and also HQAM has low PAPR than other modulation schemes due to the densest packing of constellation points. This clearly indicates the superiority and applicability for practical systems of HQAM scheme over the other QAM schemes. In Fig. 2(b), for an ASER value of $10^{-5}$, 4-SQAM has an SNR gain of 0.1 dB (approx.) over 4-HQAM which illustrates the increase in SNR gain of 4-HQAM  when compared with 2x2x2 antenna configuration. However, as the constellation size increases, HQAM performs better than the other QAM schemes. For 16 points, HQAM has an SNR gain of 0.41 dB as compared to SQAM and for 64 points,  HQAM has an SNR gain of 0.59 dB. For 32 points, HQAM has an SNR gain of 0.38 dB over 32-XQAM scheme. Since, TAS strategy is employed, where the best transmit antenna which increases the received SNR strength is choosen, the improved SNR strengths for all the QAM schemes illustrate the TAS effect in 4x4x4 antenna configuration over the 2x2x2 antenna configuration. From Fig. 2(a) and Fig. 2(b), at ASER of $10^{-5}$, the 4x4x4 antenna configuration provides the SNR gains for 4-HQAM is 6.1 dB, for 8-HQAM 6.1 dB, for 16-HQAM 6.1 dB, for 32-HQAM 6.25 dB, and for 64-HQAM 6.3 dB over the 2x2x2  antenna configuration.

\section{Conclusion}
   In this paper, the ASER expressions for the general order HQAM, general order RQAM, and 32-XQAM are derived for MIMO cooperative relay by considering the direct path with transmit antenna selection strategy. For the first time, generalized analysis and comparison of HQAM over RQAM, SQAM and 32-XQAM schemes is performed in cooperative MIMO relay network. Impact of the pathloss and the number of antennas are highlighted for different QAM schemes over the Rayleigh channels. In future, the impact of channel estimation error for  AF-MIMO systems can be investigated.

\ifCLASSOPTIONcaptionsoff
\newpage
\fi

\bibliographystyle{IEEEtran}
\small
\bibliography{TASbibfile}

\end{document}